\documentclass[preprint,journal]{vgtc}            % preprint (journal style)

\onlineid{6265}

\vgtccategory{Research}

\vgtcpapertype{algorithm/technique}

\title{Graph Drawing Stress Model with Resistance Distances}

\author{%
  \authororcid{Yosuke Onoue}{0000-0003-2739-3249}
}

\authorfooter{
  \item
  	Yosuke Onoue is with Nihon University.
  	E-mail: onoue.yousuke@nihon-u.ac.jp
}

\abstract{%
This paper challenges the convention of using graph-theoretic shortest distance in stress-based graph drawing.
We propose a new paradigm based on resistance distance, derived from the graph Laplacian's spectrum, which better captures global graph structure.
This approach overcomes theoretical and computational limitations of traditional methods, as resistance distance admits a natural isometric embedding in Euclidean space.
Our experiments demonstrate improved neighborhood preservation and cluster faithfulness.
We introduce Omega, a linear-time graph drawing algorithm that integrates a fast resistance distance embedding with random node-pair sampling for Stochastic Gradient Descent (SGD).
This comprehensive random sampling strategy, enabled by efficient pre-computation of resistance distance embeddings, is more effective and robust than pivot-based sampling used in prior algorithms, consistently achieving lower and more stable stress values.
The algorithm maintains $O(|E|)$ complexity for both weighted and unweighted graphs.
Our work establishes a connection between spectral graph theory and stress-based layouts, providing a practical and scalable solution for network visualization.
}


\teaser{
  \centering
  \includegraphics[width=\linewidth]{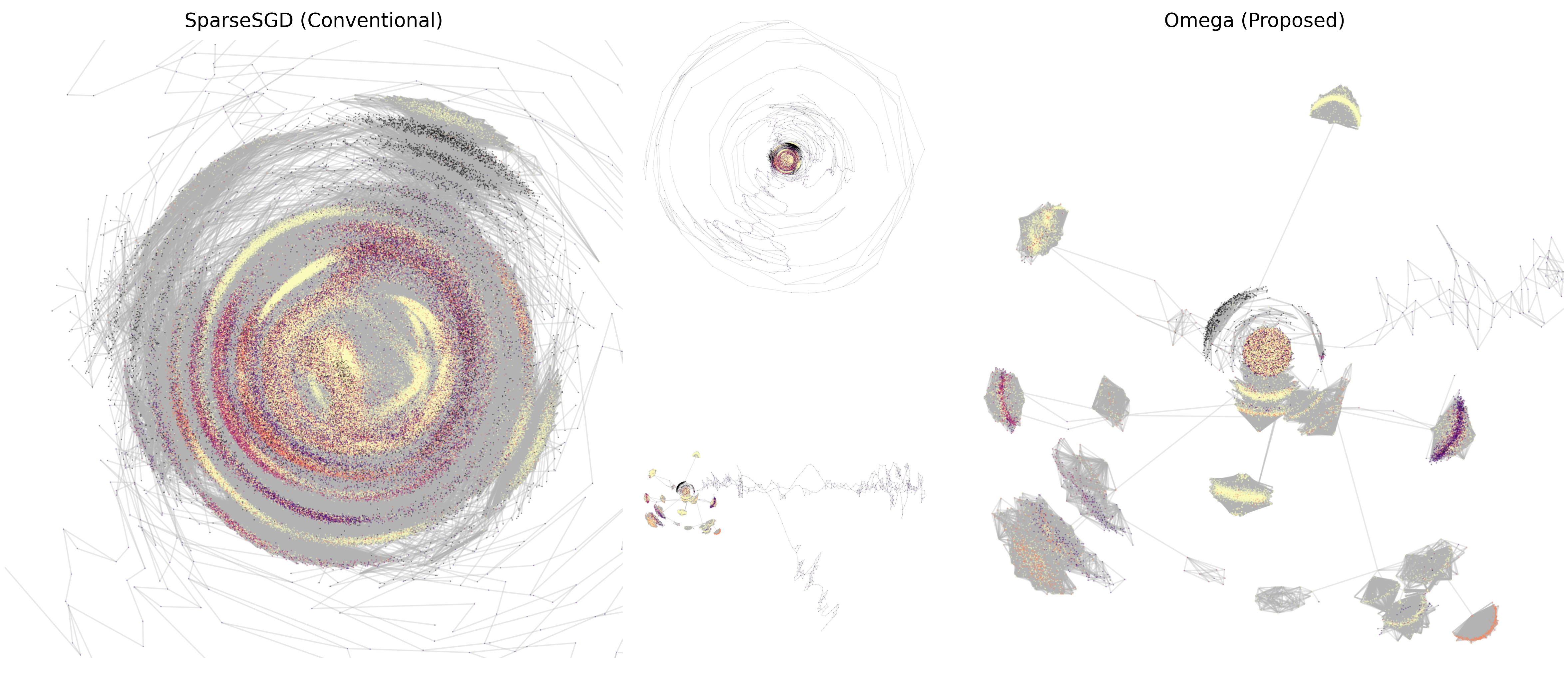}
  \caption{%
    A comparison of layouts for the web-Stanford graph ($|V|=255,265, |E|=1,941,926$) generated by SparseSGD (Conventional) and Omega (Proposed).
    Nodes are colored based on their core number.
    The conventional SparseSGD method produces a dense, unreadable "hairball," obscuring the graph's structure.
    In contrast, our proposed Omega algorithm successfully untangles the graph, revealing distinct structural components and separating the high-core nodes (center) from peripheral structures.
    The zoomed-in views highlight how Omega preserves the fine-grained local structure that is lost in the SparseSGD layout.
  }
  \label{fig:teaser}
}

\graphicspath{{figs/}{figures/}{pictures/}{images/}{./}} % where to search for the images

\usepackage{tabu}                      % only used for the table example
\usepackage{booktabs}                  % only used for the table example
\usepackage{lipsum}                    % used to generate placeholder text
\usepackage{mwe}                       % used to generate placeholder figures

\usepackage{mathptmx}                  % use matching math font

\usepackage{amsmath}
\usepackage{amssymb}
\usepackage{amsfonts}
\usepackage{algorithm}
\usepackage{algorithmicx}
\usepackage{algpseudocode}

\begin{document}

%%%%%%%%%%%%%%%%%%%%%%%%%%%%%%%%%%%%%%%%%%%%%%%%%%%%%%%%%%%%%%%%
%%%%%%%%%%%%%%%%%%%%%% START OF THE PAPER %%%%%%%%%%%%%%%%%%%%%%
%%%%%%%%%%%%%%%%%%%%%%%%%%%%%%%%%%%%%%%%%%%%%%%%%%%%%%%%%%%%%%%%

\firstsection{Introduction}

\maketitle

%% \section{Introduction} %for journal use above \firstsection{..} instead
Force-directed algorithms are a cornerstone of network visualization, producing aesthetically pleasing and structurally meaningful layouts~\cite{tamassia2013handbook}.
Stress models seek to arrange nodes such that geometric distances match ideal graph-theoretic shortest distances~\cite{kobourov2013force}.
For decades, the community has relied on graph-theoretic shortest distance as the standard.

This convention presents theoretical and computational challenges.
Graph-theoretic shortest distances may not admit an isometric embedding in Euclidean space~\cite{linial1995geometry}, meaning any layout has inherent stress.
Computing all-pairs shortest paths is expensive, particularly for large or weighted graphs.
The choice of ideal distance profoundly impacts layout quality, underscoring the need for a metric better suited for specific visual tasks~\cite{onoue2024distance}.

We propose replacing graph-theoretic shortest distance with low-rank resistance distance as the ideal metric for stress models.
Resistance distance, derived from the pseudo-inverse of the graph Laplacian~\cite{klein1993resistance}, offers key advantages.
It can be isometrically embedded in Euclidean space, eliminating inherent stress and making it compatible with stress optimization.
This spectral foundation allows resistance distance to capture global graph structure, such as clusters and bottlenecks, more effectively than path-based measures.
While shortest-path layouts excel at path-following, resistance-distance layouts produce improved neighborhood and cluster faithfulness.
Our experiments confirm clear improvements in these metrics over conventional methods.

To operationalize these benefits, we introduce Omega, a novel linear-time graph drawing algorithm built upon the SparseSGD framework~\cite{zheng2018graph}.
Omega is our name for the complete layout method that integrates three key components: (1) RDMDS for computing low-rank resistance distance embeddings, (2) a novel comprehensive random node-pair sampling strategy, and (3) SparseSGD optimization.
The low-rank structure of resistance distance allows for the efficient pre-computation of an embedding, which forms the basis of our algorithm and enables the random sampling approach.
This sampling strategy is the key algorithmic innovation that distinguishes Omega from traditional pivot-based methods, consistently achieving lower stress values than the pivot-based strategies used in prior SparseSGD implementations.
The result is an algorithm that produces higher-quality visualizations, with a favorable asymptotic complexity that is particularly advantageous for weighted graphs.

The main contributions of this work are:
\begin{itemize}
    \item A new stress model paradigm using low-rank resistance distance as the ideal distance, which yields layouts with improved neighborhood preservation and cluster faithfulness compared to traditional models.
    \item Omega, a novel linear-time graph drawing algorithm that integrates resistance distance computation with a random sampling strategy for SparseSGD. We show this sampling is more effective and robust than pivot-based methods, a benefit unlocked by the efficient, on-demand calculation of resistance distances.
    \item A demonstration of Omega's $O(|E|)$ complexity for both weighted and unweighted graphs, offering a theoretical performance advantage over existing scalable stress-based methods.
\end{itemize}

The remainder of this paper is organized as follows.
Section~\ref{sec:related} reviews related work in graph drawing, stress models, and resistance distance applications.
Section~\ref{sec:method} introduces the mathematical foundation of resistance distances and presents our stress model formulation.
Section~\ref{sec:rdmds} presents RDMDS, a linear-time algorithm for obtaining low-rank resistance distance embeddings in Euclidean space.
Section~\ref{sec:algorithm} describes the Omega algorithm in detail.
This includes the linear-time eigenvalue computation and the SparseSGD framework with resistance distances.
Section~\ref{sec:experiments} presents computational experiments.
We compare faithfulness metrics, drawing results, sampling strategies, and computational performance.
Section~\ref{sec:drawing_examples} provides a gallery of drawing examples and discusses their qualitative features.
Section~\ref{sec:discussion} discusses the theoretical implications and practical benefits of our approach.
Finally, Section~\ref{sec:conclusion} concludes the paper and outlines directions for future research.

\section{Related Work}
\label{sec:related}

Our work builds upon three main areas of research: stress-based graph drawing, spectral graph theory, and the use of resistance distance in graph algorithms.

\subsection{The Evolution of Stress Models in Graph Drawing}

Kamada and Kawai~\cite{kamada1989algorithm} pioneered "stress" in graph drawing, formulating layout as an energy minimization problem matching geometric distances to graph-theoretic shortest distances.
Gansner, Koren, and North~\cite{gansner2004graph} achieved a breakthrough with stress majorization.
Scalable methods include multilevel techniques~\cite{gansner2012maxent, meyerhenke2017drawing}, low-rank approximations~\cite{khoury2012drawing}, and sparse models.
Ortmann et al.~\cite{ortmann2017sparse} introduced the sparse stress model, optimizing over a sparse subset of node pairs.
Zheng et al.~\cite{zheng2018graph} adapted this to SGD with SparseSGD, using pivot-based sampling for million-node graphs.
Brandes and Pich~\cite{brandes2006eigensolver} proposed PivotMDS for classical MDS.

Extensions include constrained layouts~\cite{dwyer2009constrained, wang2017revisiting, dwyer2005dig} and non-Euclidean spaces~\cite{miller2022spherical, chen2020doughnets, watanabe2024cell, miller2022browser}.
However, these works predominantly use graph-theoretic shortest distance.
The choice of ideal distances has received limited discussion~\cite{onoue2024distance}.
We propose resistance distance as a theoretically grounded alternative.

\subsection{Spectral Graph Theory and Resistance Distance}

Resistance distance, first introduced by Klein and Randic~\cite{klein1993resistance}, is a metric derived from an analogy to electrical circuits, where it represents the effective resistance between two nodes.
It is formally defined for any pair of vertices and has found applications in diverse fields, from chemistry to machine learning~\cite{klein1993resistance}.
Its theoretical foundation lies in spectral graph theory~\cite{chung1997spectral}, which studies a graph's properties via the eigenvalues and eigenvectors of its Laplacian matrix.
The eigenvectors of the Laplacian are central to the definition of resistance distance and are famously used in spectral clustering to find community structure~\cite{von2007tutorial}.
Koren~\cite{koren2003spectral} also explored the direct use of Laplacian eigenvectors for spectral graph drawing, establishing a fundamental link between a graph's spectral properties and its visual representation.

From the perspective of graph signal processing (GSP)~\cite{ortega2022introduction}, our approach can be seen as a form of low-pass filtering.
By constructing our ideal distances from the low-frequency components of the graph spectrum (i.e., the eigenvectors associated with the smallest non-zero eigenvalues), our model effectively filters out high-frequency noise and preserves the smooth, global structure of the graph, which is essential for creating clear and meaningful layouts.

\subsection{Resistance Distance in Scalable Graph Drawing}

The powerful properties of resistance distance have recently been leveraged to develop scalable algorithms for force-directed graph drawing.
Eades et al.~\cite{eades2017drawing} used effective resistance for spectral graph sparsification to accelerate layout computation.
Building on this, a significant line of work by Meidiana, Hong, Eades et al.~\cite{meidiana2020sublinear, meidiana2021sublinear, meidiana2024sublinearforce} has focused on developing sublinear-time force calculation algorithms.
Their SubLinearForce algorithm~\cite{meidiana2024sublinearforce} uses effective resistance-based sampling to approximate the repulsive forces in a traditional force-directed model, achieving impressive scalability.
Notably, their work also includes an approach for sublinear-time stress minimization~\cite{meidiana2020sublinear}, where effective resistance is used to sample important node pairs for an SGD-based optimization of the classical stress model.

Our work is inspired by this research but differs in a fundamental way.
While prior work has successfully used resistance distance as a tool for sampling or approximating forces within models that still ultimately rely on graph-theoretic distances, we propose a more foundational shift: we replace the graph-theoretic shortest distance with the resistance distance as the ideal distance metric itself within the stress model (Eq.~\ref{eq:general_stress}).
This is a key distinction.
Our novelty lies not in using resistance distance in a graph drawing algorithm, but in formulating and optimizing a new stress model where resistance distance defines the ground truth.

This conceptual shift unlocks a critical algorithmic advantage.
Because our ideal distances are derived from a pre-computed low-rank embedding (RDMDS), we can calculate the ideal distance between any pair of nodes in constant time.
This liberates our algorithm from the constraints of pivot-based sampling, which is a necessary compromise in traditional SparseSGD due to the high cost of on-the-fly shortest path calculations.
Instead, Omega can employ a comprehensive random node-pair sampling strategy.
As we demonstrate in our experiments, this more uniform sampling leads to a more effective and robust minimization of the stress objective, regardless of the underlying distance metric.
Thus, our contribution is a synergistic combination: a new, theoretically-grounded stress model and a more effective sampling strategy that is enabled by it.

\section{Stress Model with Resistance Distance}
\label{sec:method}

This section details the mathematical foundation of our proposed stress model.
We first review the standard definitions for graphs and the general stress model framework.
We then introduce resistance distance, explain its theoretical underpinnings, and formulate our novel stress model based on this metric.

Let $G = (V, E)$ be an undirected, connected graph with $n = |V|$ vertices and an edge set $E$.
For weighted graphs, a function $w: E \rightarrow \mathbb{R}^+$ assigns a positive weight to each edge.
The adjacency matrix $\mathbf{A}$ and the diagonal degree matrix $\mathbf{D}$ are defined in the standard way.
The graph Laplacian matrix is then given by $\mathbf{L} = \mathbf{D} - \mathbf{A}$.
The Laplacian $\mathbf{L}$ of a connected graph is positive semi-definite.
It has exactly one zero eigenvalue, $\lambda_1=0$~\cite{von2007tutorial}.
Its eigenvalues are denoted $0 = \lambda_1 < \lambda_2 \leq \cdots \leq \lambda_n$, with corresponding eigenvectors $\mathbf{u}_1, \ldots, \mathbf{u}_n$.

A stress model finds a $d$-dimensional layout $\mathbf{X} = [\mathbf{p}_1, \ldots, \mathbf{p}_n]^T \in \mathbb{R}^{n \times d}$ minimizing:
\begin{equation}
\label{eq:general_stress}
\text{stress}(\mathbf{X}) = \sum_{i<j} w_{ij} \left(\|\mathbf{p}_i - \mathbf{p}_j\|_2 - \delta_{ij}\right)^2,
\end{equation}
where $\|\mathbf{p}_i - \mathbf{p}_j\|_2$ is the Euclidean distance in the layout, $\delta_{ij}$ is the ideal distance, and $w_{ij}$ is a corresponding weight.

A common weighting scheme, which we also adopt, is $w_{ij} = \delta_{ij}^{-2}$~\cite{kamada1989algorithm,gansner2004graph,zheng2018graph}.
This gives higher importance to preserving the distances between nearby nodes.
The optimization problem is then to find the layout $\mathbf{X}^*$ that minimizes this stress function.

Final layout quality depends critically on the choice of ideal distances $\delta_{ij}$.
Resistance distance offers superior properties compared to graph-theoretic shortest distance.
The resistance distance $r_{ij}$ between vertices $v_i$ and $v_j$ is formally defined using the Moore-Penrose pseudoinverse of the Laplacian, $\mathbf{L}^+$:
\begin{equation}
r_{ij} = (\mathbf{e}_i - \mathbf{e}_j)^T \mathbf{L}^+ (\mathbf{e}_i - \mathbf{e}_j),
\end{equation}
where $\mathbf{e}_i$ is the $i$-th standard basis vector.

This definition is best understood through its analogy to electrical circuits~\cite{klein1993resistance}.
If we imagine the graph as an electrical network where each edge is a unit resistor, $r_{ij}$ is the effective resistance between nodes $v_i$ and $v_j$.
It is the voltage difference that would be measured between $v_i$ and $v_j$ if a unit of current were injected at $v_i$ and extracted at $v_j$.
This physical intuition is powerful.
The resistance distance will be small for nodes connected by many short paths (low-resistance connections).
It will be large for nodes connected only by a few, long paths or through bottlenecks (high-resistance connections).
The Moore-Penrose pseudoinverse $\mathbf{L}^+$ arises naturally from this analogy.
It provides the unique voltage solution to Kirchhoff's current law under the constraint of a zero-mean potential, effectively grounding the circuit~\cite{von2007tutorial}.

The pseudoinverse can be expressed through the spectral decomposition of $\mathbf{L}$:
\begin{equation}
\mathbf{L}^+ = \sum_{k=2}^n \frac{1}{\lambda_k} \mathbf{u}_k \mathbf{u}_k^T.
\end{equation}
This leads to a formulation of resistance distance based on the eigenvalues and eigenvectors of the Laplacian:
\begin{equation}
r_{ij} = \sum_{k=2}^n \frac{(u_{k,i} - u_{k,j})^2}{\lambda_k},
\end{equation}
where $u_{k,i}$ is the $i$-th component of eigenvector $\mathbf{u}_k$.

\begin{figure*}[tb]
  \centering
  \includegraphics[width=\linewidth]{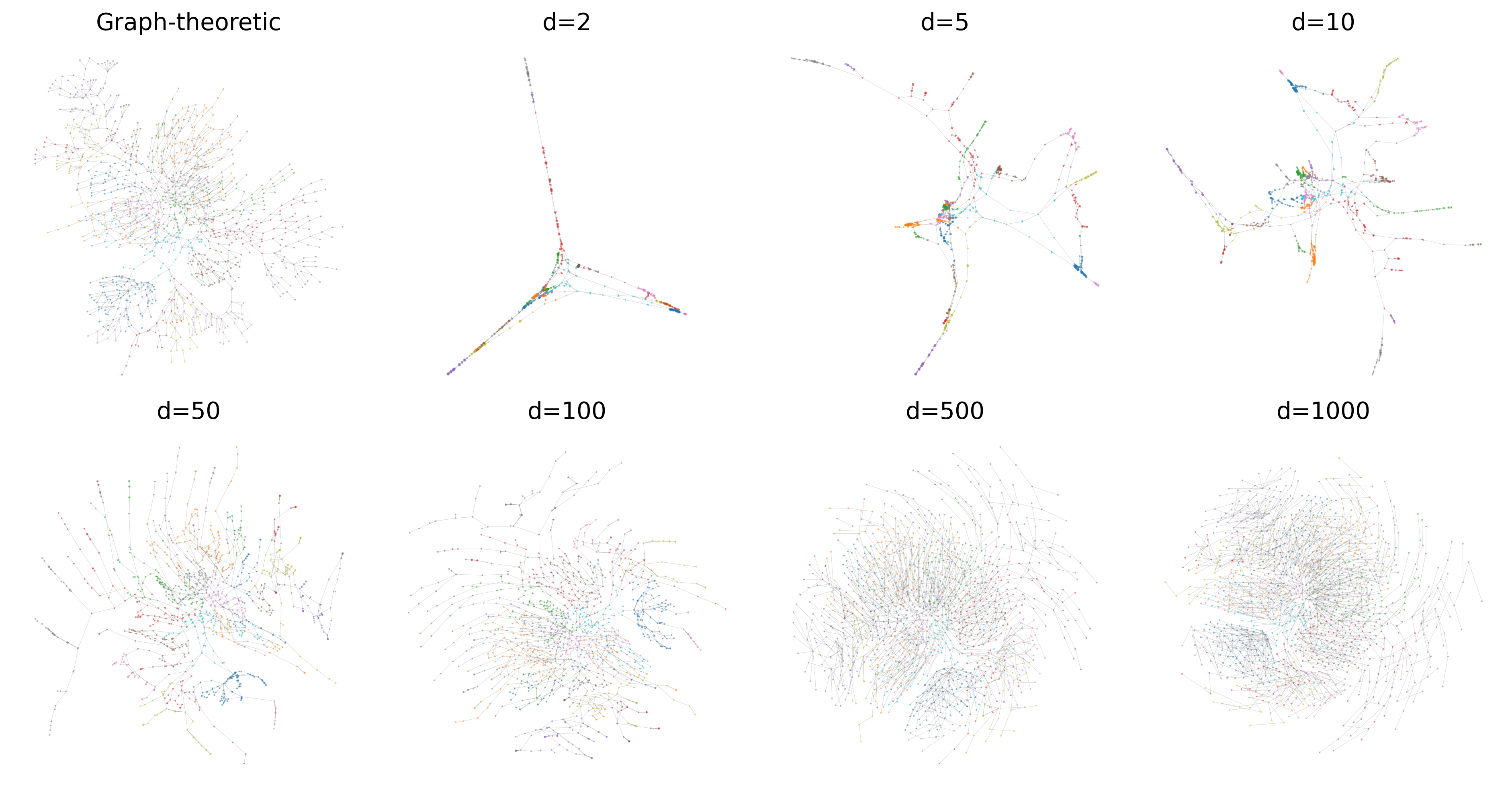}
  \caption{
    Visual comparison of layouts for the `1138\_bus` graph ($|V|=1138, |E|=1458$) from the SuiteSparse Matrix Collection~\cite{davis2011university}.
    Each panel shows the final 2D layout generated by optimizing a stress model with a different ideal distance metric.
    The top-left panel uses the conventional graph-theoretic shortest distance.
    The subsequent panels use low-rank resistance distance, where the rank $d$ of the embedding used to calculate the ideal distances is varied.
    All layouts were generated by applying FullSGD (1000 iterations) followed by Stress Majorization until convergence, with nodes colored by community.
    This demonstrates the expressiveness of the metric: small ranks ($d=5, 10, 50$) clearly separate communities by capturing global structure, while very low ranks ($d=2$) lose local detail.
    Sufficiently large ranks ($d=100, 500, 1000$) reflect most of the graph's information but can cause peripheral branches to curl inwards, an effect of the high-dimensional Euclidean embedding.
  }
  \label{fig:low-rank-comparison}
\end{figure*}

A key insight for computational efficiency is to use a low-rank approximation.
For a given rank $d \ll n$, we define the low-rank resistance distance by summing over only the first $d$ non-zero eigenvalues:
\begin{equation}
\tilde{r}_{ij}^{(d)} = \sum_{k=2}^{d+1} \frac{(u_{k,i} - u_{k,j})^2}{\lambda_k}.
\end{equation}
This approximation captures the dominant spectral components of the graph, which correspond to its global structure.
\Cref{fig:low-rank-comparison} visually demonstrates the effect of varying the rank $d$ on the final layout.

Crucially, this low-rank resistance distance can be embedded isometrically in a $d$-dimensional Euclidean space.
By defining a set of embedding coordinates $\mathbf{X} \in \mathbb{R}^{n \times d}$ for each vertex $v_i$ as
\begin{equation}
\mathbf{X}_{i,:} = \left(\frac{u_{2,i}}{\sqrt{\lambda_2}}, \frac{u_{3,i}}{\sqrt{\lambda_3}}, \ldots, \frac{u_{d+1,i}}{\sqrt{\lambda_{d+1}}}\right)^T,
\end{equation}
the Euclidean distance between any two points $\|\mathbf{X}_{i,:} - \mathbf{X}_{j,:}\|_2$ is exactly equal to the square root of the low-rank resistance distance, $\sqrt{\tilde{r}_{ij}^{(d)}}$:
\begin{equation}
\label{eq:isometric_embedding}
\|\mathbf{X}_{i,:} - \mathbf{X}_{j,:}\|_2 = \sqrt{\tilde{r}_{ij}^{(d)}}.
\end{equation}
This isometric embedding property is fundamental to our approach.
It means that there exists a geometric configuration in a $d$-dimensional space whose Euclidean distances perfectly match the square root of the resistance distances.
This is a key theoretical advantage over graph-theoretic shortest distances, which generally cannot be embedded without distortion and thus have inherent stress.
It is important to note that this $d$-dimensional embedding $\mathbf{X}$ is used only to compute the ideal distances; the final layout is a separate 2D entity, $\mathbf{Y}$, that is optimized to approximate these ideal distances.

We therefore formulate our stress model by using the general stress function (Equation~\ref{eq:general_stress}) with ideal distances $\delta_{ij}$ based on the square root of the low-rank resistance distances.
A potential issue arises for automorphic nodes, which will have identical embedding coordinates, resulting in an ideal distance of zero.
To prevent this, we enforce a minimum distance $\varepsilon_d$, defining the ideal distance as:
\begin{equation}
\delta_{ij} = \max\left(\sqrt{\tilde{r}_{ij}^{(d)}}, \varepsilon_d \right).
\end{equation}
This approach offers several advantages over traditional models.
These include a natural and distortion-free low-dimensional embedding, a smooth distance function that captures global structure, and efficient computation via spectral methods.

The results in \Cref{fig:low-rank-comparison} illustrate the practical impact of varying the rank $d$.
For small ranks, such as $d=5, 10,$ and $50$, the layout effectively captures the global structure of the graph, leading to a clear separation of the communities identified by community detection.
When the rank is too low, as in the case of $d=2$, much of the local information is lost, resulting in a layout where the detailed structure is less discernible.
Conversely, for sufficiently large ranks ($d=100, 500, 1000$), the layout incorporates almost all of the graph's structural information.
However, this can lead to a tendency for the outer branches to curl inwards.
This phenomenon occurs because resistance distance can reduce the distinction between the distances of peripheral nodes and those closer to the core, a direct consequence of the properties of high-dimensional Euclidean embeddings.
This trade-off highlights that selecting an appropriately small rank $d$ is key to creating layouts that best reflect the graph's overall structure without introducing artifacts from the embedding process.

\section{RDMDS: Resistance Distance MDS}
\label{sec:rdmds}

To operationalize our stress model, we require an efficient method for computing the low-rank resistance distance embedding.
This section introduces RDMDS (Resistance Distance Multidimensional Scaling).
RDMDS is a linear-time algorithm designed for this purpose.
It computes the coordinates $\mathbf{x}_i$ for each vertex by finding the $d$ smallest non-zero eigenvalues and corresponding eigenvectors of the graph Laplacian $\mathbf{L}$.
Crucially, it achieves this in $O(|E|)$ time, assuming the target dimension $d$ is a small constant.

The core of RDMDS is the orthogonal inverse power method~\cite{golub2013matrix}.
While existing high-performance eigensolvers are available, we present a self-contained implementation to prioritize portability and provide a lightweight solution tailored for this research context.
This iterative algorithm is well-suited for finding the smallest eigenvalues of large, sparse symmetric matrices like the graph Laplacian.
To find a single eigenvector, the standard inverse power method iteratively solves a linear system of the form $(\mathbf{L} + \sigma \mathbf{I})\mathbf{v}^{(k+1)} = \mathbf{v}^{(k)}$.
Here, $\sigma > 0$ is a small shift that makes the matrix invertible without significantly altering the eigenvectors.
To find multiple orthogonal eigenvectors, we extend this with a deflation technique based on Gram-Schmidt orthogonalization~\cite{strang2012linear}.
After computing each eigenvector, we project it out from the iteration vector to ensure the method converges to a new, orthogonal eigenvector.
By enforcing a fixed maximum number of iterations, we guarantee that this process remains computationally bounded.

Each step of the inverse power method requires solving a large linear system.
To improve convergence, we use the Preconditioned Conjugate Gradient (PCG) method for this task.
It is ideal for solving systems involving symmetric positive-definite matrices like our shifted Laplacian.
As a preconditioner, we use the Incomplete Cholesky factorization with zero fill-in (IC(0))~\cite{saad2003iterative}.
For a sparse matrix $\mathbf{A}$, the preconditioner $\mathbf{K}$ can be computed in $O(|E|)$ time, and solving the system $\mathbf{Kz} = \mathbf{r}$ is also an $O(|E|)$ operation.
The PCG method leverages the sparsity of the Laplacian matrix, performing matrix-vector products in $O(|E|)$ time.
By setting a constant maximum number of iterations for the PCG solver, each linear system is solved in $O(|E|)$ time.

We now formally prove the $O(|E|)$ complexity of RDMDS.
It is important to note that conventional algorithms for computing exact resistance distances typically have near-linear time complexity \cite{spielman2014nearly}.
However, for use as ideal distances in stress models, high numerical precision is not essential.
Our RDMDS algorithm exploits this observation by computing a low-rank approximation with controlled iteration counts, achieving true linear time $O(|E|)$ complexity at the cost of approximate rather than exact resistance distances.
This trade-off is well-suited for graph drawing applications where the goal is visual clarity rather than numerical exactness.

The algorithm's computational cost is dominated by the eigenvalue computation, which consists of nested iterations.
We analyze each component:

\begin{enumerate}
    \item \textbf{Preconditioner construction:} Computing the IC(0) factorization of the sparse matrix $\mathbf{A}$ requires $O(|E|)$ time, as it processes only the non-zero entries of the Laplacian.
    
    \item \textbf{PCG iteration complexity:} Each PCG iteration involves:
    \begin{itemize}
        \item Matrix-vector multiplication $\mathbf{Ap}$: $O(|E|)$ for a sparse matrix
        \item Preconditioner solve $\mathbf{Kz} = \mathbf{r}$: $O(|E|)$ using IC(0)
        \item Vector operations (dot products, additions): $O(|V|) \subseteq O(|E|)$
    \end{itemize}
    Thus, each PCG iteration costs $O(|E|)$.
    
    \item \textbf{Linear system solve:} Solving $(\mathbf{L} + \sigma\mathbf{I})\mathbf{v}^{(k+1)} = \mathbf{v}^{(k)}$ requires at most $M_{\text{CG}}$ PCG iterations, yielding $O(M_{\text{CG}} \cdot |E|)$ complexity per solve.
    
    \item \textbf{Inverse power iteration:} Each inverse power iteration performs one linear system solve plus Gram-Schmidt orthogonalization against $k-1$ previous eigenvectors. The orthogonalization costs $O(k|V|) \subseteq O(d|V|) \subseteq O(|E|)$ for constant $d$ in connected graphs where $|E| \geq |V|-1$. Combined with the linear solve, each iteration costs $O(M_{\text{CG}} \cdot |E|)$.
    
    \item \textbf{Finding $d$ eigenvectors:} For each of $d$ eigenvectors, we perform at most $M_{\text{eig}}$ inverse power iterations, yielding $O(d \cdot M_{\text{eig}} \cdot M_{\text{CG}} \cdot |E|)$ total complexity.
\end{enumerate}

For constant parameters $d$, $M_{\text{eig}}$, and $M_{\text{CG}}$, the overall complexity simplifies to $O(|E|)$.
This linear-time performance is essential for the scalability of our overall approach.

The RDMDS algorithm has a deep connection to classical spectral graph drawing~\cite{koren2003spectral}.
When $d=2$, the embedding produced by RDMDS is equivalent to the layout from an aspect-ratio balanced spectral drawing.
In spectral drawing, node positions are simply the eigenvectors corresponding to the smallest non-zero eigenvalues.
The aspect-ratio balanced variant scales these eigenvectors by the inverse square root of their corresponding eigenvalues.
This is precisely what RDMDS computes.
This establishes that our model, in its unweighted form, is a generalization of classical spectral methods.
It provides a bridge between the two major paradigms of stress-based and spectral layouts.

The embedding coordinates $\mathbf{x}_i$ produced by RDMDS are highly versatile.
They allow for the efficient computation of the ideal distances for our stress model, as the Euclidean distance between embedding coordinates corresponds to the square root of the resistance distance: $\|\mathbf{x}_i - \mathbf{x}_j\|_2 = \sqrt{\tilde{r}_{ij}^{(d)}}$.
This avoids the need to store a full distance matrix.
For graph drawing ($d=2$), these coordinates serve as an excellent initial layout.
They already capture the global structure of the graph.
For larger $d$, RDMDS acts as a principled dimensionality reduction technique, preserving resistance distance relationships.
These coordinates are not only the optimal solution for an unweighted resistance-distance stress model.
They are also an ideal starting point for the weighted optimization we describe next.

Algorithm~\ref{alg:rdmds} presents the complete RDMDS procedure.
The algorithm takes as input a connected graph $G$ and target dimension $d$, and outputs the embedding coordinates $\mathbf{X} \in \mathbb{R}^{n \times d}$ for all vertices.

\begin{algorithm}[t]
\caption{RDMDS: Resistance Distance Multidimensional Scaling}
\label{alg:rdmds}
\begin{algorithmic}[1]
\Require Graph $G = (V, E)$ with $n = |V|$ vertices, target dimension $d$
\Ensure Embedding coordinates $\mathbf{X} \in \mathbb{R}^{n \times d}$
\State \textbf{Parameters:} shift parameter $\sigma$, eigenvalue tolerance $\tau_{\text{eig}}$, max eigenvalue iterations $M_{\text{eig}}$, CG tolerance $\tau_{\text{CG}}$, max CG iterations $M_{\text{CG}}$
\State Compute Laplacian matrix $\mathbf{L}$ from graph $G$
\State $\mathbf{A} \leftarrow \mathbf{L} + \sigma \mathbf{I}$ \Comment{Shift to make invertible}
\State $\mathbf{K} \leftarrow \text{IncompleteCholesky}(\mathbf{A})$ \Comment{Preconditioner for PCG}
\State $\mathbf{X} \leftarrow \mathbf{0}_{n \times d}$
\For{$k = 1$ to $d$} \Comment{Compute $d$ smallest eigenvectors}
    \State $\mathbf{v} \leftarrow \text{random vector}$
    \For{$j = 1$ to $k-1$} \Comment{Gram-Schmidt orthogonalization}
        \State $\mathbf{v} \leftarrow \mathbf{v} - (\mathbf{v}^T \mathbf{X}_{:,j}) \mathbf{X}_{:,j}$
    \EndFor
    \State $\mathbf{v} \leftarrow \mathbf{v} / \|\mathbf{v}\|_2$
    \For{$\text{iter} = 1$ to $M_{\text{eig}}$} \Comment{Inverse power iteration}
        \State $\mathbf{v}_{\text{old}} \leftarrow \mathbf{v}$
        \State $\mathbf{v} \leftarrow \text{PCG}(\mathbf{A}, \mathbf{v}, \mathbf{K}, \tau_{\text{CG}}, M_{\text{CG}})$
        \For{$j = 1$ to $k-1$} \Comment{Deflation step}
            \State $\mathbf{v} \leftarrow \mathbf{v} - (\mathbf{v}^T \mathbf{X}_{:,j}) \mathbf{X}_{:,j}$
        \EndFor
        \State $\mathbf{v} \leftarrow \mathbf{v} / \|\mathbf{v}\|_2$
        \If{$|\mathbf{v}^T \mathbf{L} \mathbf{v} - \mathbf{v}_{\text{old}}^T \mathbf{L} \mathbf{v}_{\text{old}}| < \tau_{\text{eig}}$}
            \State \textbf{break}
        \EndIf
    \EndFor
    \State $\lambda_k \leftarrow \mathbf{v}^T \mathbf{L} \mathbf{v} - \sigma$ \Comment{Rayleigh quotient}
    \State $\mathbf{X}_{:,k} \leftarrow \mathbf{v} / \sqrt{\lambda_k}$
\EndFor
\State \Return $\mathbf{X}$
\end{algorithmic}
\end{algorithm}

The algorithm requires the auxiliary procedure for solving linear systems, as detailed in Algorithm~\ref{alg:cg}.

\begin{algorithm}[t]
\caption{Preconditioned Conjugate Gradient (PCG)}
\label{alg:cg}
\begin{algorithmic}[1]
\Require Matrix $\mathbf{A}$, right-hand side $\mathbf{b}$, preconditioner $\mathbf{K}$, tolerance $\tau_{\text{CG}}$, maximum iterations $M_{\text{CG}}$
\Ensure Solution $\mathbf{x}$ to $\mathbf{A}\mathbf{x} = \mathbf{b}$
\State $\mathbf{x} \leftarrow \mathbf{0}$
\State $\mathbf{r} \leftarrow \mathbf{b}$
\State Solve $\mathbf{Kz} = \mathbf{r}$ for $\mathbf{z}$ \Comment{Apply preconditioner}
\State $\mathbf{p} \leftarrow \mathbf{z}$
\State $\rho_{\text{old}} \leftarrow \mathbf{r}^T \mathbf{z}$
\For{$\text{iter} = 1$ to $M_{\text{CG}}$} \Comment{Iterative refinement}
    \State $\mathbf{q} \leftarrow \mathbf{A} \mathbf{p}$
    \State $\alpha \leftarrow \rho_{\text{old}} / (\mathbf{p}^T \mathbf{q})$
    \State $\mathbf{x} \leftarrow \mathbf{x} + \alpha \mathbf{p}$
    \State $\mathbf{r} \leftarrow \mathbf{r} - \alpha \mathbf{q}$
    \State Solve $\mathbf{Kz} = \mathbf{r}$ for $\mathbf{z}$ \Comment{Apply preconditioner}
    \State $\rho_{\text{new}} \leftarrow \mathbf{r}^T \mathbf{z}$
    \If{$\rho_{\text{new}} < \tau_{\text{CG}}^2 $} \Comment{Check convergence}
        \State \textbf{break}
    \EndIf
    \State $\beta \leftarrow \rho_{\text{new}} / \rho_{\text{old}}$
    \State $\mathbf{p} \leftarrow \mathbf{z} + \beta \mathbf{p}$ \Comment{Update search direction}
    \State $\rho_{\text{old}} \leftarrow \rho_{\text{new}}$
\EndFor
\State \Return $\mathbf{x}$
\end{algorithmic}
\end{algorithm}

The RDMDS algorithm provides a robust and efficient foundation for computing resistance distance embeddings, enabling the development of scalable algorithms for large-scale graph visualization and analysis.

\section{Omega Layout Algorithm}
\label{sec:algorithm}

This section introduces Omega, our novel linear-time graph drawing algorithm.
The name, derived from the symbol for electrical resistance ($\Omega$), reflects its foundation in resistance distance.
Omega builds upon the SparseSGD framework, incorporating several key innovations that are unlocked by the use of resistance distance.

The most significant innovation is a shift away from pivot-based sampling, the standard for scalable stress models.
Traditional SparseSGD implementations use pivot-based sampling because computing graph-theoretic shortest paths for arbitrary node pairs on the fly is too expensive.
This reliance on a small number of pivots can bias the optimization and lead to suboptimal layouts.
Omega overcomes this limitation.
By pre-computing the RDMDS embedding, we can calculate the ideal resistance distance for any pair of nodes in constant time.
This enables a more comprehensive and effective random node-pair sampling strategy, leading to better stress minimization.

Omega also holds a critical computational advantage: uniform complexity.
Traditional SparseSGD is faster on unweighted graphs (using BFS) than on weighted graphs (requiring Dijkstra's algorithm).
Omega's complexity, however, remains $O(|E|)$ for both weighted and unweighted graphs.
This is because its RDMDS pre-computation is insensitive to edge weights.
Furthermore, the core sparse matrix operations in RDMDS are highly amenable to GPU acceleration.
This offers a path to even greater performance gains.

The Omega algorithm follows a six-step procedure that integrates the RDMDS embedding with SGD optimization.
\begin{enumerate}
    \item \textbf{Compute Embedding $\mathbf{X}$:} First, we use RDMDS to compute the $d$-dimensional low-rank resistance distance embedding $\mathbf{X}$ for all vertices. This one-time, $O(|E|)$ pre-computation is the foundation for the subsequent steps.
    \item \textbf{Initialize Pair Set:} We create an initial set of node pairs $P$ by including all edges from the graph. This ensures the local structure is explicitly considered.
    \item \textbf{Random Sampling:} For each vertex, we randomly sample $h$ additional node pairs to add to $P$. This is the key step that distinguishes Omega from pivot-based methods, allowing for a more uniform and comprehensive sampling of the distance constraints.
    \item \textbf{Compute Ideal Distances $\delta_{ij}$:} For every pair $(i, j)$ in $P$, we compute the ideal distance $\delta_{ij} = \|\mathbf{X}_{i,:} - \mathbf{X}_{j,:}\|_2$ from the pre-computed embedding. Thanks to the pre-computed embedding, this is a fast, constant-time operation for each pair.
    \item \textbf{Initialize 2D Layout $\mathbf{Y}$:} We use the first two dimensions of the RDMDS embedding $\mathbf{X}$ as the initial 2D layout coordinates $\mathbf{Y}$. This provides an excellent starting point that already reflects the graph's global structure.
    \item \textbf{SGD Optimization:} Finally, we use an iterative SGD process with an annealing schedule to refine the 2D layout $\mathbf{Y}$. At each step, we iterate through a shuffled list of the pairs in $P$, updating the node positions to minimize the stress function with respect to the ideal distances $\delta_{ij}$.
\end{enumerate}

The overall computational complexity of Omega is $O(|E|)$ for a constant sampling parameter $h$.
The RDMDS pre-computation is $O(|E|)$.
The sampling and distance computation steps create and process $O(|E| + h|V|) = O(|E|)$ pairs.
The final SGD optimization consists of a fixed number of passes over these pairs.
This combination of a theoretically sound distance metric and an efficient, scalable algorithm makes Omega a powerful tool for large-scale network visualization.

Omega was implemented as part of our graph drawing package, \texttt{Egraph}\footnote{\url{https://github.com/likr/egraph-rs}}.
Egraph provides implementations of major stress-based graph drawing algorithms, including Stress Majorization and SGD.
The core of Egraph is implemented in Rust for high-speed execution.
It also provides interfaces for Python and JavaScript, enabling flexible network visualization in scripting languages.

Algorithm~\ref{alg:omega} presents the complete Omega procedure.
The algorithm takes as input a connected graph $G$, target dimensions $d$ for the resistance distance embedding and $2$ for the final layout, sampling parameter $h$, and various optimization parameters for the SGD process.

\begin{algorithm}[t]
\caption{Omega: Resistance Distance Graph Drawing}
\label{alg:omega}
\begin{algorithmic}[1]
\Require Graph $G = (V, E)$, embedding dimension $d$, sampling parameter $h$, min ideal distance $\varepsilon_d$
\Ensure 2D layout coordinates $\mathbf{Y} \in \mathbb{R}^{n \times 2}$
\State \textbf{Parameters:} step size $\eta$, convergence tolerance $\tau$
\State $\mathbf{X} \leftarrow \text{RDMDS}(G, d)$ \Comment{Compute resistance distance embedding}
\State $P \leftarrow \{(i, j) \mid (v_i, v_j) \in E\}$
\For{$i = 1$ to $n$} \Comment{Random node-pair sampling}
    \For{$l = 1$ to $h$}
        \State $j \leftarrow \text{random}(1, n)$
        \If{$i \neq j$ and $(i, j) \notin P$}
            \State $P \leftarrow P \cup \{(i, j)\}$
        \EndIf
    \EndFor
\EndFor
\For{$(i, j) \in P$} \Comment{Compute ideal distances and weights}
    \State $\delta_{ij} \leftarrow \max(\|\mathbf{X}_{i} - \mathbf{X}_{j}\|_2, \varepsilon_d)$
    \State $w_{ij} \leftarrow \delta_{ij}^{-2}$
\EndFor
\State $\mathbf{Y} \leftarrow \mathbf{X}_{:,1:2}$ \Comment{Initialize 2D layout}
\For{$\tau$ in annealing schedule} \Comment{SGD optimization}
    \For{$(i, j) \in P$ in random order}
        \State $\mu \leftarrow \min(1, w_{ij} \tau)$
        \State $\mathbf{r} \leftarrow \frac{\|\mathbf{Y}_{i} - \mathbf{Y}_{j}\| - \delta_{ij}}{2} \frac{\mathbf{Y}_{i} - \mathbf{Y}_{j}}{\|\mathbf{Y}_{i} - \mathbf{Y}_{j}\|}$
        \State $\mathbf{Y}_{i} \leftarrow \mathbf{Y}_{i} - \mu \mathbf{r}$
        \State $\mathbf{Y}_{j} \leftarrow \mathbf{Y}_{j} + \mu \mathbf{r}$
    \EndFor
\EndFor
\State \Return $\mathbf{Y}$
\end{algorithmic}
\end{algorithm}

The algorithm integrates seamlessly with the RDMDS procedure from Section~\ref{sec:rdmds}.
It uses the computed embedding coordinates as both the source of resistance distance values and the initial layout positions.
The annealing schedule controls the step size throughout the optimization process.
It begins with large updates that allow for global rearrangements and gradually reduces to fine-scale adjustments.
The random shuffling of vertex pairs at each iteration ensures that the optimization process explores different aspects of the stress function, leading to more robust convergence.

Omega represents a significant advancement in stress-based graph drawing algorithms.
It combines the theoretical advantages of resistance distances with practical computational efficiency.
The algorithm's linear time complexity, uniform performance across graph types, and improved sampling strategy make it particularly suitable for large-scale graph visualization applications.
In these applications, both quality and computational efficiency are important considerations.

\section{Computational Experiments}
\label{sec:experiments}

To validate our proposed approach, we conducted a series of computational experiments designed to evaluate the performance of our resistance distance-based stress model and the Omega algorithm.
This section presents the results of these experiments, which are organized into three parts.
First, we compare the faithfulness of layouts generated using our proposed low-rank resistance distance against the traditional graph-theoretic shortest distance.
Second, we evaluate the effectiveness of Omega's random node-pair sampling strategy against the pivot-based sampling used in traditional SparseSGD.
Finally, we analyze the computational performance and scalability of Omega.

All experiments were conducted on a machine with an AMD Ryzen 5 5600X 6-Core Processor (3.70 GHz) and 64 GB of RAM.

\subsection{Comparison of Faithfulness Metrics}

We first compare the quality of layouts generated by a standard stress model using two different ideal distance metrics: the traditional graph-theoretic shortest distance (the baseline) and our proposed low-rank resistance distance for various embedding dimensions $d$.
To isolate the effect of the distance metric itself, we used a consistent optimization setting for both.
Specifically, we applied FullSGD, which uses all node pairs rather than a sparse sample, for 15 iterations.
This experimental design serves a different purpose than the scalable algorithm comparison in Section 6.2: here, we aim to compare how different distance metrics perform under standard parameter settings, evaluating their practical effectiveness rather than theoretical optimality.
The FullSGD component processes all node pairs (not just a sparse sample), ensuring that the comparison reflects the inherent properties of each distance metric.
The 15-iteration setting represents a typical practical configuration for FullSGD optimization.
For the resistance distance metric, a minimum ideal distance of $\varepsilon_d=0.01$ was enforced.
The experiments were conducted on a comprehensive dataset of 213 graphs from the SuiteSparse Matrix Collection~\cite{davis2011university}, selecting graphs with node counts ranging from 100 to 1,000, with results for each graph being the median of 10 trials.

We use two primary metrics to quantify layout faithfulness, assessing how well a layout preserves the intrinsic structure of the graph~\cite{eades2017shape, nguyen2017proxy,cai2024cluster}.
\begin{itemize}
    \item \textbf{Neighborhood Preservation:} We measure the average Jaccard coefficient between the k-nearest neighbors in the original graph and the k-nearest neighbors in the resulting layout for each node~\cite{eades2017shape,nguyen2017proxy}.
    \item \textbf{Clustering Quality:} Following the method of Cai et al.~\cite{cai2024cluster}, we first apply greedy modularity community detection to the input graph to establish ground-truth clusters. We then perform agglomerative clustering on the layout coordinates to obtain the same number of clusters and compute the Fowlkes-Mallows score~\cite{fowlkes1983method} between the two clusterings.
\end{itemize}

\begin{figure}[tb]
  \centering
  \includegraphics[width=\linewidth]{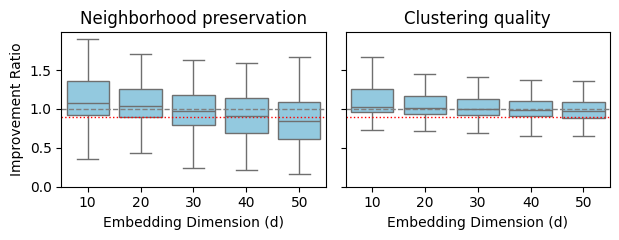}
  \caption{Comparison of faithfulness metrics. For each metric, we show the improvement ratio, calculated as (Proposed Metric Score) / (Baseline Metric Score), for our method with different embedding dimensions ($d=10, 20, \dots, 50$) over the baseline. For both metrics, a ratio greater than 1.0 indicates an improvement.}
  \label{fig:quality_comparison}
\end{figure}

\Cref{fig:quality_comparison} shows the improvement ratio of using resistance distance over the baseline.
For both metrics, resistance distance demonstrates superior performance, with the median ratio consistently exceeding 1.0.
This strong performance is not an empirical artifact but an expected outcome rooted in spectral graph theory.
Faithfulness metrics, by design, evaluate how well a layout preserves the intrinsic high-dimensional structure of a graph, such as its neighborhood and community structures.
Resistance distance is fundamentally defined by the graph Laplacian's spectrum, and it is well-established that the low-frequency components of this spectrum (i.e., the eigenvectors associated with the smallest non-zero eigenvalues) encode precisely this global structure.
Therefore, a distance metric built upon these components is theoretically predisposed to excel at preserving the very qualities these metrics measure.

Notably, the method achieves strong performance even at a low dimension of $d=10$.
This suggests that a compact, low-rank approximation is sufficient to capture the most critical structural information.
The slight decrease in quality as $d$ increases further indicates that higher-frequency spectral components may introduce noise rather than meaningful structural detail, reinforcing the effectiveness of a low-rank approach.
The risk of degradation is also low; even in the worst-performing cases, the first quartile of the improvement ratio remains high.
This confirms that the potential for significant, theoretically-grounded improvement far outweighs the risk of minor degradation.

\subsection{Comparison of Sampling Strategies}
\label{sec:sampling_strategies}

To evaluate the effectiveness of our proposed random node-pair sampling strategy, we compared it against the traditional pivot-based sampling commonly used in SparseSGD implementations.
This experiment focuses on the performance of scalable algorithms, so we used SparseSGD with 15 iterations as the optimization engine for all conditions, which is a practical setting for large graphs.
We used the same comprehensive dataset described in Section 6.1.
For each graph, we ran both sampling strategies using both graph-theoretic and resistance distances as the underlying metric.
The number of samples per vertex, $h$, was varied from 10 to 100.
To ensure robust results, each experimental condition was repeated 10 times with different random seeds.
When using our random sampling strategy with resistance distance, we set the embedding dimension to $d=10$ and the minimum ideal distance to $\varepsilon_d=0.01$.
The performance was measured by the ratio of the final stress value achieved by the algorithm to the known optimal stress value for the given distance metric.
For the baseline optimal stress value, we applied FullSGD for 1000 iterations followed by Stress Majorization until convergence.
Since obtaining the true global optimum of the stress model is theoretically difficult, we use this well-converged result as the reference optimal solution.
This allows us to compare how close each scalable sampling strategy gets to the theoretical best.

\begin{figure}[t]
  \centering
  \includegraphics[width=\linewidth]{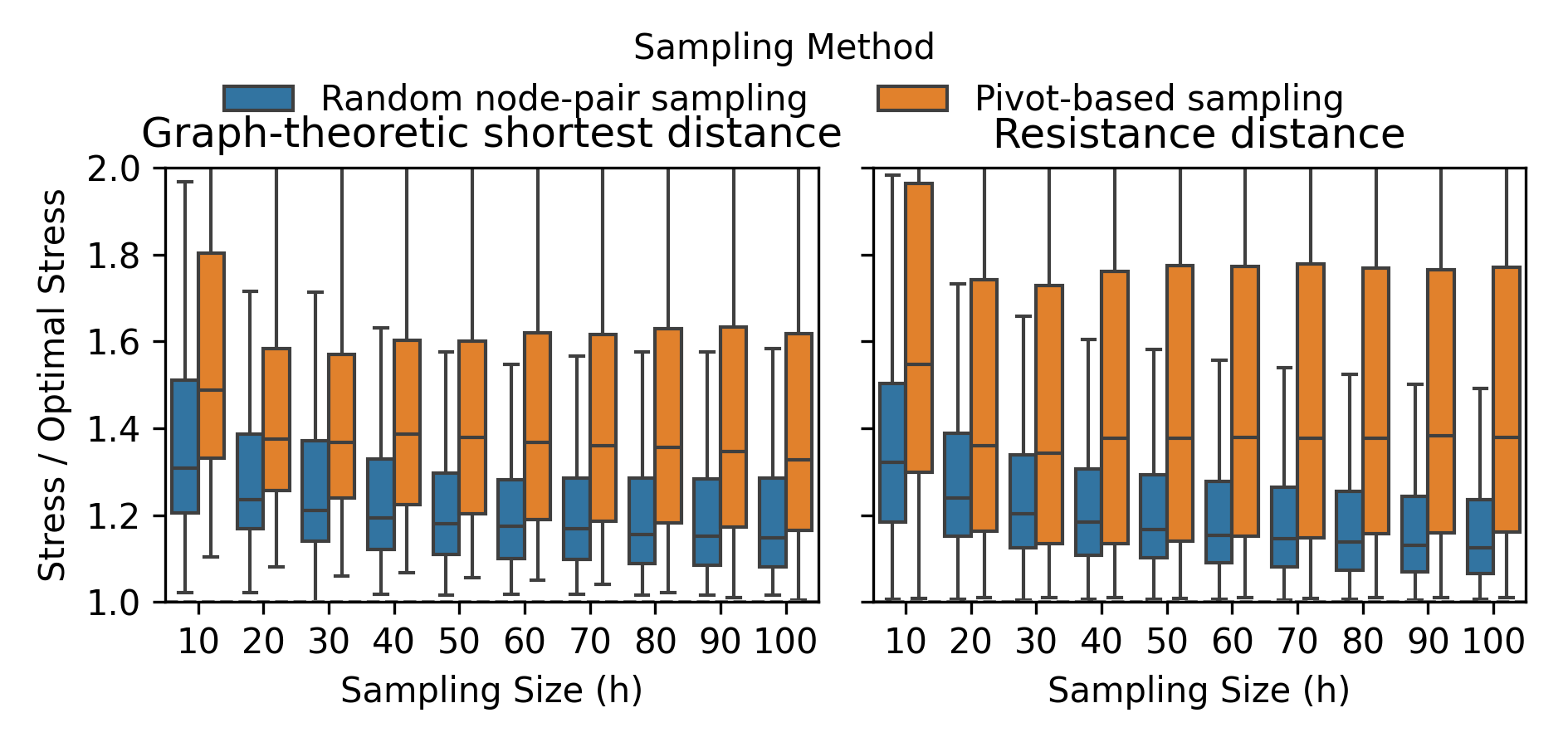}
  \caption{
    Comparison of stress ratios for Random Node-Pair Sampling (our Omega algorithm) and Pivot-Based Sampling.
    The y-axis shows the ratio of the achieved stress to the optimal stress (a value of 1.0 is optimal).
    The box plots summarize the distribution of ratios over our entire graph dataset for each sampling size $h$.
    (Top) Using the graph-theoretic shortest distance, our method (blue) consistently outperforms pivot-based sampling (orange).
    (Bottom) Using resistance distance, the superiority of our method is also clearly demonstrated.
  }
  \label{fig:sampling_comparison}
\end{figure}

\Cref{fig:sampling_comparison} presents a detailed comparison of the final stress ratios achieved by our random node-pair sampling strategy versus a traditional pivot-based approach.
The results unequivocally demonstrate the superiority of our method across all tested conditions.

For both distance metrics, the boxes corresponding to our random sampling (blue) are consistently and significantly lower than those for pivot-based sampling (orange).
This indicates that our method achieves stress values much closer to the theoretical optimum.
Notably, the median stress ratio (the line within each blue box) is always substantially lower, and in many cases, the entire interquartile range (the box itself) for our method lies below the median of the pivot-based method.
This highlights a robust performance advantage.

Furthermore, the variance in performance is dramatically lower with our approach.
The blue boxes are consistently tighter, and their whiskers are shorter, signifying that our random sampling strategy yields more stable and reliable results across the diverse set of graphs in our test suite.

A key finding of this experiment is the robustness of our random sampling strategy.
While the performance gains are most pronounced when paired with resistance distance, our method also demonstrates a clear and consistent advantage for the traditional graph-theoretic shortest distance metric.
This indicates that the core benefit of our approach—the move from a biased, pivot-based sample to a comprehensive, random one—is a fundamental improvement, independent of the specific distance metric used.
The ability to efficiently pre-compute an embedding (such as RDMDS for resistance distance) is what unlocks this superior sampling strategy.
This provides the SGD optimizer with a more uniform and unbiased view of the global stress landscape, allowing it to find significantly better solutions and reinforcing the idea that sampling is a critical component of the optimization's success.

\subsection{Computational Time Comparison}

Finally, we analyze the computational performance and scalability of Omega.
The experiment was conducted on 947 real-world matrix data from the SuiteSparse Matrix Collection, which have between 100 and 1,000,000 edges and fewer than 1,000,000 non-zero elements, by graphing them and extracting the largest connected components.
For SparseSGD's shortest path search, Dijkstra's algorithm was used.
The number of iterations for each SGD was set to 15, with an $\epsilon$ of 0.1 and a sampling number h of 50.
For Omega, the embedding dimension d was 10, the minimum ideal distance was set to $\varepsilon_d=0.01$, the maximum number of iterations for the inverse power method was 100, and the maximum number of iterations for the conjugate gradient method was 100.
For each condition, 10 trials were conducted with different random number seeds, and the median of the computation times was taken.

\begin{figure}[tb]
  \centering
  \includegraphics[width=\linewidth]{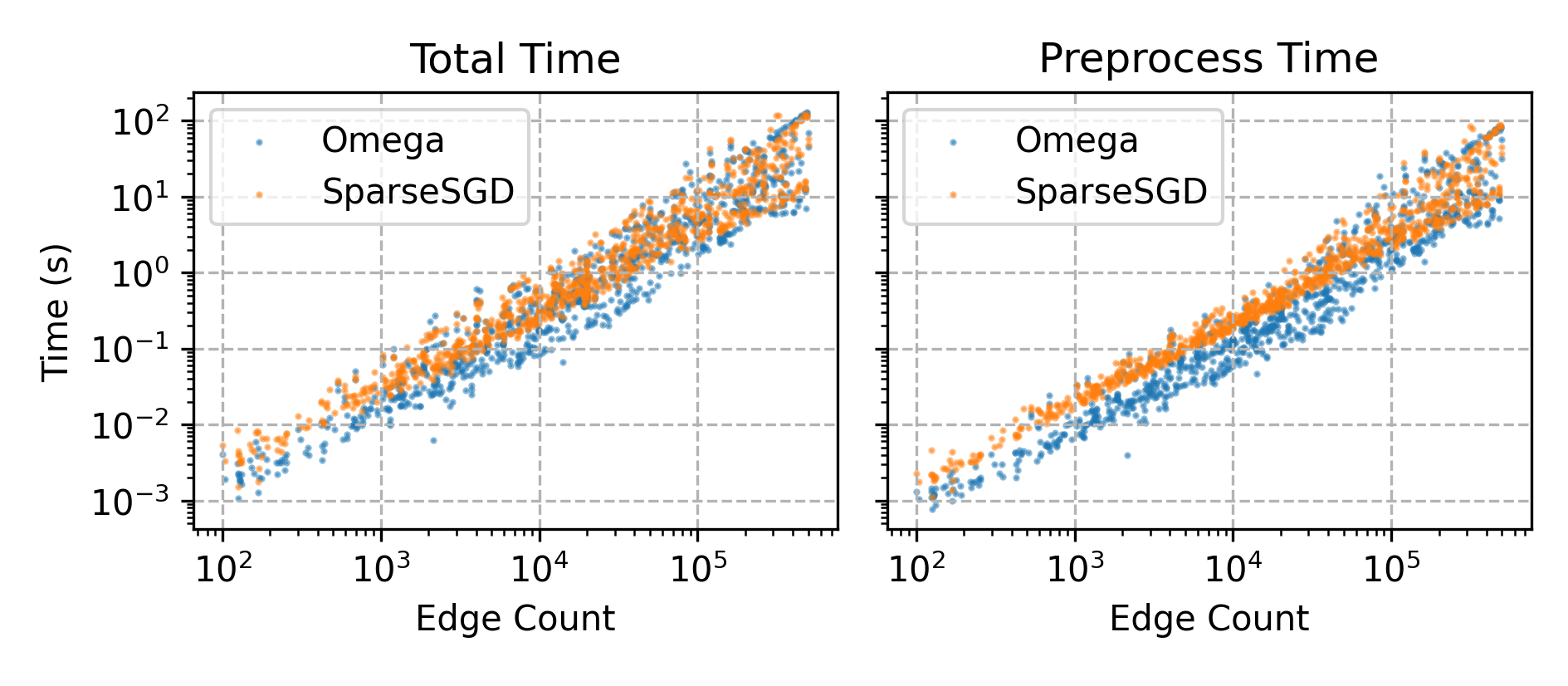}
  \caption{Comparison of computational time. The top and bottom panels show the total time and the preprocessing time, respectively. The horizontal axis is the number of edges, and the vertical axis is the computation time. Both axes are on a log scale. To avoid perceptual bias from overplotting, points are rendered with transparency.}
  \label{fig:runtime_comparison}
\end{figure}

\Cref{fig:runtime_comparison} shows the relationship between the number of edges and the computation time.
Theoretically, Omega holds an asymptotic advantage for weighted graphs, with its $O(|E|)$ complexity being superior to the $O(|E| + |V|\log|V|)$ of SparseSGD's Dijkstra-based implementation.
However, the experimental results indicate that this theoretical benefit does not always translate to faster runtimes in practice.
The majority of the computation time for both methods is occupied by the preprocessing step (distance calculation and pair list creation).
While Omega's preprocessing is often faster for small to medium-sized graphs, its runtime for larger graphs (approximately $|E| > 10^5$) can be longer.
This is primarily because the constant factor in its complexity is larger, driven by the number of iterations required for the conjugate gradient solver in RDMDS to converge.
Therefore, while Omega offers a more favorable theoretical scaling, its practical performance is comparable to SparseSGD.
The linear time complexity of RDMDS is contingent on the number of PCG iterations being constant, but in practice, this number can grow for larger or more complex graphs, increasing the constant factor in the overall complexity.
Future improvements in preconditioning or the use of GPU acceleration could help mitigate this and realize the theoretical advantage more consistently.

\section{Drawing Examples}
\label{sec:drawing_examples}

We begin our qualitative comparison with the teaser figure (\Cref{fig:teaser}), which provides a striking example of Omega's ability to handle large, complex graphs like `web-Stanford` from the SuiteSparse Matrix Collection~\cite{davis2011university}.
The nodes are colored based on their core number~\cite{batagelj2003m}.
The layout produced by the conventional SparseSGD method is a dense, unreadable "hairball," a common failure mode for large graphs where the global structure is lost in a tangle of edges.
In stark contrast, Omega's layout is clear and informative.
It successfully untangles the core of the graph from its periphery, revealing distinct structural components that were previously hidden.
This is a direct result of using resistance distance, which captures the global relationships between nodes more effectively than shortest-path distance.
By preserving the separation between structurally distant parts of the graph, Omega produces a layout that is not only aesthetically superior but also provides genuine insight into the network's organization.
The zoomed-in views further confirm that this global clarity does not come at the cost of local detail; fine-grained structures are well-preserved.

\begin{figure*}[tb]
  \centering
  \includegraphics[width=\linewidth]{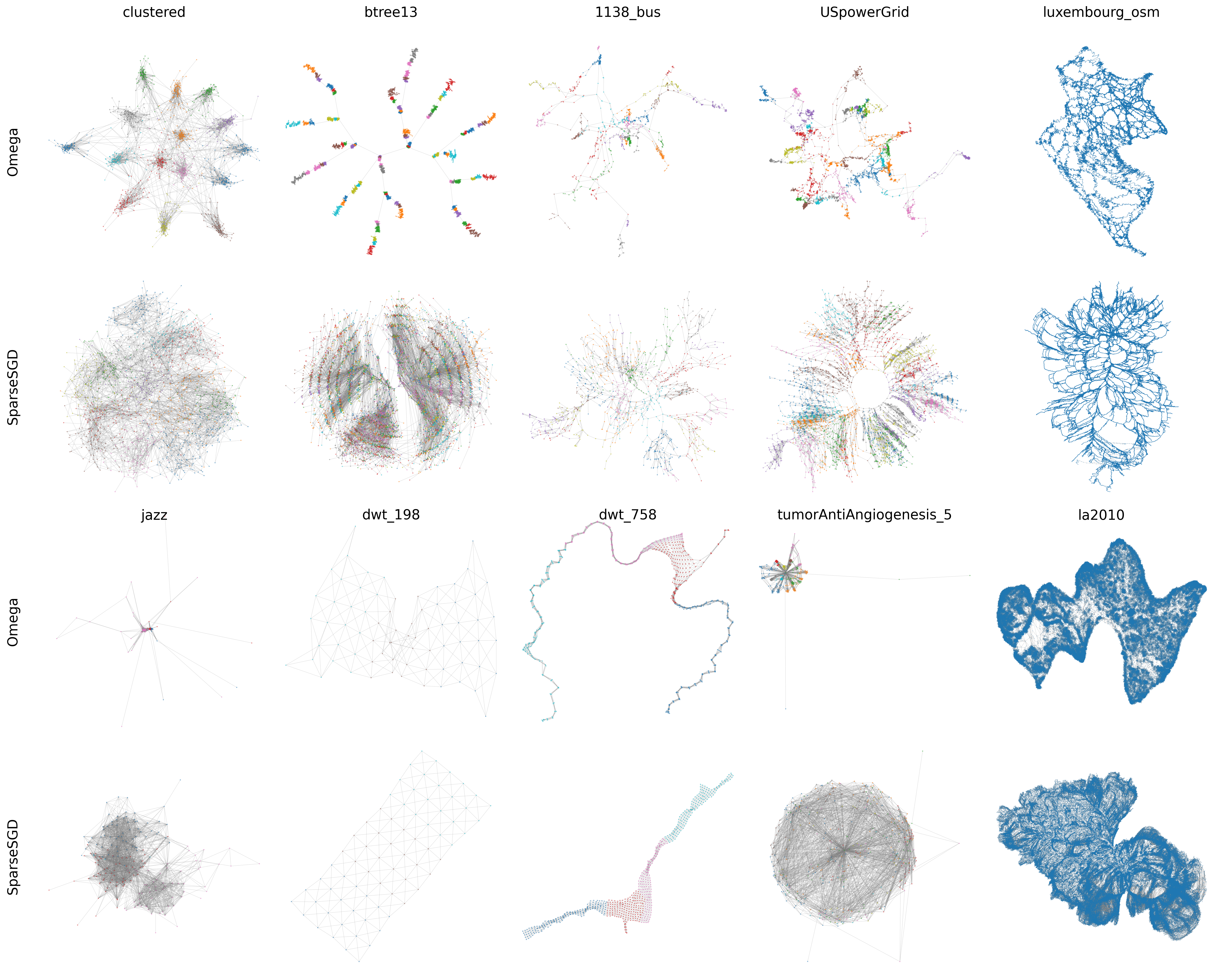}
  \caption{
    A gallery of layouts for 10 representative graphs, comparing the output of Omega (top row of each pair) and SparseSGD (bottom row of each pair).
    Omega consistently produces layouts that better articulate cluster structures and global relationships.
  }
  \label{fig:comparison_grid}
\end{figure*}

\begin{table}[tb]
  \centering
  \caption{Statistics for the graphs used in the drawing examples.}
  \label{tab:graph_stats}
  \begin{tabular}{lrr}
    \toprule
    Graph & $|V|$ & $|E|$ \\
    \midrule
    web-Stanford & 255,265 & 1,941,926 \\
    clustered & 1,495 & 3,964 \\
    btree13 & 8,192 & 8,191 \\
    1138\_bus & 1,138 & 1,458 \\
    USpowerGrid & 4,941 & 6,594 \\
    luxembourg\_osm & 114,599 & 119,666 \\
    jazz & 198 & 2,742 \\
    dwt\_198 & 72 & 236 \\
    dwt\_758 & 758 & 2,618 \\
    tumorAntiAngiogenesis\_5 & 460 & 2,207 \\
    la2010 & 204,447 & 490,317 \\
    \bottomrule
  \end{tabular}
\end{table}

\Cref{fig:comparison_grid} presents a gallery of layouts for 10 representative graphs, comparing the output of Omega and SparseSGD.
The statistics for these graphs are summarized in \Cref{tab:graph_stats}.
The `clustered` graph is a Random Partition Graph~\cite{fortunato2010community} generated using NetworkX, consisting of 15 clusters of 100 nodes each; the intra-cluster and inter-cluster edge probabilities were set to 0.05 and 0.0002, respectively.
The `btree13` graph is a binomial tree of order 13.
The remaining graphs were obtained from the SuiteSparse Matrix Collection~\cite{davis2011university}.
For all graphs except `luxembourg\_osm` and `la2010`, the node colors represent communities detected using the method of Clauset et al.~\cite{clauset2004finding}.
All layouts were generated with 15 SGD iterations, an $\epsilon$ of 0.1, and a sampling parameter $h=50$.
For the Omega layouts, we additionally set the embedding dimension to $d=10$ and the minimum ideal distance to $\varepsilon_d=0.01$.

In the `clustered`, `btree13`, `1138\_bus`, and `USpowerGrid` graphs, the node coloring clearly shows that Omega generates layouts that properly separate the clusters.
Notably, for the `1138\_bus` graph, the layout produced by our approximate Omega algorithm is comparable to the high-precision result in \Cref{fig:low-rank-comparison}, which was generated using FullSGD.
The graphs `jazz`, `dwt\_198`, `dwt\_758`, and `tumorAntiAngiogenesis\_5` are representative examples that showed extreme results in the faithfulness metrics comparison.

The `jazz` graph had a clustering quality improvement ratio of 0.74.
This is likely because Omega's layout clearly separates a dense core of nodes from peripheral nodes, which may have led to inaccurate cluster identification by the metric's algorithm.
The `dwt\_198` graph had a clustering quality improvement ratio of 0.73.
This is attributed to the graph's uniform grid structure; the low-rank resistance distance may not have sufficiently reflected the local grid structure.
Conversely, `dwt\_758`, a similar grid-like graph, showed the highest neighborhood preservation improvement ratio of 2.83.
In this case, the low-rank resistance distance successfully captured the global grid structure, leading to a significant improvement.

The `tumorAntiAngiogenesis\_5` graph exhibited the largest improvement in clustering quality (2.43) but also the most significant degradation in neighborhood preservation (0.24).
This graph has a central node connected to most other nodes, which causes the "hairball" problem in the SparseSGD layout.
Omega resolves this by keeping intra-cluster distances short, thus achieving proper cluster separation.
However, its inability to maintain proximity to the central node is thought to be the cause of the worsened neighborhood preservation.

For `luxembourg\_osm`, SparseSGD produces a layout that bulges outwards.
In contrast, Omega folds the long path-like subgraphs in a zigzag pattern, suppressing the outward expansion while maintaining the relative positions of the clusters.
In the `la2010` graph, Omega not only suppresses outward expansion similarly to `luxembourg\_osm` but also successfully draws local node clusters without distortion.

\section{Discussion}
\label{sec:discussion}

Our experiments provide strong evidence for the benefits of using resistance distance in stress-based graph drawing.

A key takeaway is choosing a distance metric suited to the intended visual analysis task.
Layouts based on graph-theoretic shortest distance are intuitive for path-following and estimating shortest distances between nodes.
Resistance distance, however, reveals global structure like clusters and bottlenecks.
The layouts are better suited for understanding high-level network organization.
We do not claim resistance distance is universally superior to graph-theoretic shortest distance, but rather more effective for specific visual analysis tasks.
The ideal choice depends on the user's goal.

The improved performance on faithfulness metrics is a direct consequence of the spectral properties of resistance distance, not an incidental benefit.
A key finding is the strong alignment between our distance metric and the qualities being measured.
Faithfulness metrics like neighborhood preservation and clustering quality are designed to assess how well a layout represents the graph's global structure.
This global structure is mathematically encoded in the low-frequency spectrum of the graph Laplacian.
The eigenvectors corresponding to the smallest non-zero eigenvalues are known to be smooth over the graph, varying slowly and thus partitioning the graph into its main constituent clusters~\cite{von2007tutorial}.
This is precisely the principle behind spectral clustering.
Our approach leverages this exact insight.
By defining our ideal distances using a low-rank approximation based on these very eigenvectors, we are creating a stress model that is explicitly optimized to preserve the graph's most significant structural features.
The high scores on faithfulness metrics are therefore a direct confirmation of this theoretical alignment.
The layout naturally respects the cluster structure because the distances themselves are defined by it.
This provides a clear theoretical justification for our empirical results and deepens the connection between spectral graph theory and the qualitative evaluation of graph drawings.

Furthermore, our work highlights the profound impact of choosing an appropriate "ideal distance" in stress models.
The fact that simply replacing the distance metric can lead to significant improvements in layout quality, as demonstrated in our experiments, underscores that the design of ideal distances is a critical, and perhaps under-explored, aspect of graph drawing.
Resistance distance, with its strong theoretical grounding in electrical circuit theory and spectral analysis, presents a compelling alternative to the more heuristically motivated graph-theoretic shortest distance.

The connection between our approach and classical spectral methods is also noteworthy.
As established in Section~\ref{sec:rdmds}, the solution to an unweighted stress model using low-rank resistance distance (which is equivalent to classical MDS on resistance distances) coincides with the layout produced by classical spectral graph drawing when $d=2$.
This provides a unifying bridge between two major paradigms in graph drawing: stress-based optimization and spectral methods.
Our framework can thus be seen as a generalization of spectral drawing, extending it to weighted stress models and enabling further refinement through SGD.

Finally, the computational advantages of Omega are significant.
While traditional SparseSGD using the graph-theoretic shortest distance has a complexity of $O(|E| + |V| \log |V|)$ for weighted graphs (due to Dijkstra's algorithm), Omega maintains a consistent $O(|E|)$ complexity.
This is because the RDMDS pre-computation, which depends on the graph's spectral properties, is insensitive to edge weights.
This positions Omega as a promising scalable solution for the large, weighted networks that are common in many real-world applications.

Regarding parameter selection, our computational experiments demonstrate that Omega is robust to a wide range of parameter choices.
The algorithm performs well across different values of the embedding rank $d$ and sampling parameter $h$, though certain defaults provide a good balance between quality and efficiency.
For layout quality, we found that an embedding dimension of $d=10$, a sampling parameter of $h=50$, and a minimum ideal distance of $\varepsilon_d=0.01$ work well in most cases.
Performance remains stable for $d \ge 10$, confirming that a low rank is sufficient to capture essential structural information, while the benefits of increasing $h$ beyond 50 begin to diminish.
The overall runtime is primarily dependent on the parameters for RDMDS.
The eigenvalue tolerance $\tau_{\text{eig}}$ and shift parameter $\sigma$ must be smaller than the smallest non-zero eigenvalue; for graphs with up to one million edges, values of $10^{-5}$ and $10^{-6}$, respectively, were generally effective.
We set the CG tolerance $\tau_{\text{CG}}$ to $0.1$, which typically resulted in fewer than 10 inverse power iterations and fewer than 100 CG iterations for graphs of this size.
The number of SGD iterations and the scheduler's $\epsilon$ parameter functioned well when set to values similar to those used in conventional SGD.

\section{Conclusion and Future Work}
\label{sec:conclusion}

This paper replaced graph-theoretic shortest distance with resistance distance in stress-based models.
We showed this shift overcomes limitations of traditional approaches.
Our resistance distance-based stress model produces layouts with improved faithfulness to graph structure.
We introduced Omega, a linear-time algorithm combining efficient resistance distance embedding with comprehensive random sampling for SGD.
Experiments confirm Omega produces higher-quality layouts with performance advantages for weighted graphs.
Its random sampling strategy achieves lower and more stable stress values than pivot-based sampling across different distance metrics.

This work bridges spectral graph theory and stress-based optimization, opening avenues for future research.
Omega's core computations are amenable to GPU acceleration for further speedups.
Developing adaptive methods for selecting optimal rank $d$ could refine the performance-quality trade-off.
Future work includes comparing with cluster-faithful methods like LinLog and tsNET, and exploring theoretical connections between resistance distance and clustering approaches.
Our random sampling strategy could generalize to other graph drawing and optimization problems where efficient distance pre-computation is feasible.
The principles could extend beyond graph drawing to network embedding and manifold learning.

\bibliographystyle{abbrv-doi-hyperref}

\bibliography{document}

\end{document}